\documentclass[12pt,preprint]{aastex}

\shorttitle{BUBBLES IN COOLING FLOW CLUSTERS}
\shortauthors{SOKER, BLANTON, \& SARAZIN}

\slugcomment{Submitted to the Astrophysical Journal}

\begin{document}

\title{Hot Bubbles in Cooling Flow Clusters}

\author{Noam Soker\altaffilmark{1},
Elizabeth L. Blanton\altaffilmark{2},
and Craig L. Sarazin}

\affil{Department of Astronomy, University of Virginia, P.O. Box 3818
Charlottesville, VA 22903-0818;
soker@physics.technion.ac.il, eblanton@virginia.edu,  sarazin@virginia.edu}

\altaffiltext{1}{On sabbatical from the University of Haifa at Oranim,
Department of Physics, Oranim, Tivon 36006, Israel}

\altaffiltext{2}{{\it Chandra} Postdoctoral Fellow}

\begin{abstract}
As more cooling flow clusters of galaxies with central radio sources are
observed with the {\it Chandra} and {\it XMM-Newton} X-ray Observatories,
more examples of ``bubbles'' (low-emission regions in the X-ray coincident
with radio emission) are being found.
These bubbles are surrounded by bright shells of
X-ray emission, and no evidence of current strong shocks has yet been found.
Using an analytic approach and some simplifying assumptions, we derive 
expressions relating the size and
location of a bubble, as well as the density contrast between the bubble and
the ambient medium, with the shock history of the bubble.  These can be
applied straightforwardly to new observations.  We find that existing
observations are consistent with a mild shock occurring in the past, and with
the bulk of the cool material in the X-ray shells being cooled at the cluster
center and then pushed outward by the radio source.  Strong shocks are ruled
out unless they occurred more than 1 Gyr ago.  We also discuss 
Rayleigh-Taylor instabilities as well as the case of a bubble expanding 
into an older bubble produced from a previous cycle of radio activity.
\end{abstract}


\section{Introduction} \label{sec:intro}

{\it Chandra} high spatial resolution observations of clusters of galaxies 
reveal the presence of X-ray-deficient bubbles in the inner regions of many
cooling flow clusters
(Perseus [Abell~426], Fabian et al.\ 2000, 2002;
Hydra A [Abell~780], McNamara et al.\ 2000;
Abell~2052, Blanton et al.\ 2001, hereafter BSMW; 
Abell~496, Dupke \& White 2001;
Abell~2199, Fabian 2002; 
MKW3s, Mazzotta et al.\ 2001; 
Abell~2597, McNamara et al.\ 2001;
RBS797, Schindler et al.\ 2001;
Abell~85, Kempner, Sarazin, \& Ricker 2002;
Abell~133, Fujita et al.\ 2002;
Abell~4059, Heinz et al.\ 2002).
Earlier {\it ROSAT} observations also showed a few similar results,
albeit with less resolution
(Perseus [Abell~426], B\"ohringer et al.\ 1993;
Abell~4059, Huang \& Sarazin 1998).
These bubbles are characterized by low X-ray emissivity implying low density. 
In most cases, the bubbles are sites of strong radio emission.
There are clusters where the bubbles are less well-defined, although
there may be hints to their existence  
(Abell~1795, Fabian et al.\ 2001; 
3C295, Allen et al.\ 2001b;
Virgo M87, Belsole et al.\ 2001).
The absence of evidence for shocks suggests that the bubbles are 
expanding and moving at subsonic or mildly transonic velocities
(Fabian et al.\ 2000;
McNamara et al.\ 2000;
BSMW).
The presence of bubbles which do not coincide with strong radio emission
(known as `ghost bubbles' or `ghost cavities')
located farther from the centers of the clusters
in Perseus (Fabian et al.\ 2000), 
MKW3s (Mazzotta et al.\ 2001), and Abell~2597 (McNamara et al.\ 2001),
suggests that the bubbles rise buoyantly.

The discovery of bubbles and their detailed observational study 
has stimulated theoretical studies of X-ray bubble formation and evolution
(Churazov et al.\ 2001; Nulsen et al.\ 2002, hereafter N2002; 
Fabian et al.\ 2002).  
N2002 study the origin of the cool gas in the rims of enhanced X-ray emission
in Hydra A, and particularly address the role of magnetic fields. 
Although the current paper addresses some similar questions to those
addressed by N2002, our analysis and results have only a small overlap
with those of N2002. 
Our discussion is not specific to an individual cluster.
We aim to provide simple analytical expressions that can be used
for different conditions and evolutionary stages in a variety of
cooling flow clusters.
We illustrate our results by applying them to several specific clusters.
A number of recent papers provide detailed numerical simulations of bubbles
(Rizza et al.\ 2000; 
Churazov et al.\ 2001;
Br\"uggen et al.\ 2002;
Quilis, Bower, \& Balogh 2001; 
Saxton, Sutherland, \& Bicknell 2001;
Reynolds, Heinz, \& Begelman 2001; 
Br\"uggen \& Kaiser 2001).
However, it is difficult to generalize the results of these simulations
and apply them to other observed cooling flow clusters.

In \S~\ref{sec:shock}, we briefly discuss the thermal evolution of shocked gas 
near the cluster center.
Simple expressions for the properties of a bubble inflated by hot plasma 
(e.g., from a radio jet) are derived in \S~\ref{sec:bubbles}.
In \S~\ref{sec:rt}, we show that a bubble is stable to Rayleigh-Taylor
modes at early stages, becoming unstable only at late stages and
in the outer (away from the cluster center) portion.   
Some properties of ``ghost'' bubbles (i.e., 
bubbles at relatively large radii with weak or no radio 
emission) are discussed in \S~\ref{sec:ghost}.
In \S~\ref{sec:interaction}, we consider the interaction between bubbles,
which we speculate may explain the X-ray and radio structure in the eastern 
radio `ear' in M87. 
We discuss and summarize our main results in \S~\ref{sec:discussion}.

\section{Shock Compression and Cooling} \label{sec:shock}

In several of the radio bubbles which have been observed with
{\it Chandra}, the bubbles are surrounded by relatively thin shells of dense,
cool, X-ray emitting thermal gas.
The best cases are probably Perseus (Fabian et al.\ 2000, 2002)
and Abell~2052 (BSMW).
What mechanism has produced these geometrically thin, cool, and dense
shells?
Below, we will argue that the cool, dense gas may have come from
farther in towards the center of the cooling flow.
However, simply slowly lifting (since there is no indication for strong shocks)
dense material from the inner regions will not result in thin shells and 
filaments having a sharp boundary with their ambient medium, but rather 
a continuous shallow density gradient. 
Instead, we examine the compression of the shell by a mild shock. 
For example, weak shock heating has been seen in the gas near the nucleus 
of the galaxy NGC~4636 which seems to have been
heated by a shock expanding from the nucleus (Jones et al.\ 2002).
We will assume that the origin of the shock is energy deposited by a central
AGN, probably through the action of a jet.
We will assume that gas shocks, and undergoes subsequent expansion as
it equilibrates with the ambient gas pressure, and rises from the
cooling flow center to regions of lower gas pressure.
At the same time, the gas cools radiatively.

For simplicity, we will assume a planar shock, and a $\gamma=5/3$ gas
(neglecting any magnetic field pressure),
so that the density and pressure enhancement factors at the shock are
\begin{equation}
\eta \equiv \frac{\rho_1}{\rho_i} = 
\frac{4 {\cal M}^2}{{\cal M}^2+3} \, , 
\label{eq:shock compression}
\end{equation}
and 
\begin{equation}
\alpha \equiv \frac{P_1}{P_i} = 
\frac{5 {\cal M}^2 -1}{4} \, , 
\label{eq:shock pressure}
\end{equation}
respectively.
Here, $\rho$ is the gas mass density, $P$ is the pressure, ${\cal M}$ is
the Mach number of the shock, and
the subscripts `$i$' and $`1'$ stand for initial and 
post-shock quantities, respectively, calculated at the initial location at
which the gas is shocked, not its present location.
The radiative cooling rate, which is assumed to be dominated by thermal
bremsstrahlung, is taken to be 
\begin{equation}
L=\kappa \rho^2 T^{1/2} \, .
\label{eq:cooling}
\end{equation}
We define the initial cooling time as
\begin{equation}
\tau_c \equiv \frac{5}{2} \frac {P_i}{L_i} \, .
\label{eq:cooling time}
\end{equation}
The post shock gas cools radiatively, and its entropy decreases according
to 
\begin{equation}
\frac {3}{2} P \frac{d}{dt} \ln ( P \rho^{-5/3} ) = - L \, .
\label{eq:entropy}
\end{equation}
We will assume that a time $t_1$ has transpired since the gas was
shocked, and that the gas has been cooling radiatively for this period.

After being shocked, the gas pressure drops continuously as the gas
adjusts to the local ambient pressure, and then decreases further as
the gas bubble rises out of the cooling flow center into regions of lower
ambient pressure.
Let $P_a \le P_i$ be the final ambient pressure of the gas.
A variety of different time histories for the gas pressure 
might be possible.
We consider two extreme cases, which maximize or minimize the role of
radiative cooling.

We first consider a case which maximizes the importance of radiative
cooling.
We assume that the post-shock gas is cooling radiatively at a constant
pressure $P_1=\alpha P_i$ for the entire time $t_1$, and then undergoes
a very rapid adiabatic expansion to its present pressure $P_a$
on a very short timescale, during which no radiative cooling occurs.
This gives the maximum possible cooling, since the gas spends most of
its time at the highest density and temperature values, 
where radiative cooling is most efficient. 
This maximal cooling manifests a simple analytical solution of
equation~(\ref{eq:entropy}).
The density at the end of time $t_1$, just before the adiabatic expansion, 
is
\begin{equation}
\rho_{1f} = \left(1- 
\frac{3}{2} \frac {\eta^{3/2}}{\alpha^{1/2}} 
\frac{t_1}{\tau_c} \right)^{-2/3} \rho_1 \, . 
\label{eq:max_cool_1}
\end{equation}
Note that the gas will have cooled completely if $t_1 \ge \tau_{\rm min}$,
where the minimum cooling time is
\begin{equation}
\tau_{\rm min} \equiv \frac{2}{3} \frac{\alpha^{1/2}}{\eta^{3/2}} \tau_c
\, .
\label{eq:tmin}
\end{equation}
After the rapid adiabatic cooling (during which we neglect radiative cooling),
the ratio of the final shell density $\rho_f$ to that
of its present ambient medium $\rho_a$ is 
\begin{equation}
\frac {\rho_{f}}{\rho_a} = \left( \frac{\eta}{\alpha^{3/5}} \right)
\left(1- \frac{3}{2} \frac {\eta^{3/2}}{\alpha^{1/2}} 
\frac{t_1}{\tau_c} \right)^{-2/3}
\left( \frac{\rho_i}{\rho_a} \right)
\left( \frac{P_i}{P_a} \right)^{-3/5} \, .
\label{eq:max_cool_f}
\end{equation}

The opposite extreme, which minimizes the effect of cooling, occurs
if rapid adiabatic expansion occurs just after the gas is shocked,
and then the gas cools radiatively at a pressure of $P_a$.
In this case, the gas will cool completely if $t_1 \ge \tau_{\rm max}$,
where
\begin{equation}
\tau_{\rm max} \equiv \frac{2}{3} \frac{\alpha^{9/10}}{\eta^{3/2}}
\left( \frac{P_i}{P_a} \right)^{2/5} \tau_c \, ,
\label{eq:tmax}
\end{equation}
and the ratio of the final shell density to the present ambient density is
\begin{equation}
\frac {\rho_{f}}{\rho_a} = \left( \frac{\eta}{\alpha^{3/5}} \right)
\left[ 1 - \frac{3}{2} \frac {\eta^{3/2}}{\alpha^{9/10}}
\left( \frac{P_i}{P_a} \right)^{-2/5}
\frac{t_1}{\tau_c} \right]^{-2/3}
\left( \frac{\rho_i}{\rho_a} \right)
\left( \frac{P_i}{P_a} \right)^{-3/5} \, .
\label{eq:min_cool_f}
\end{equation}

As an example, we consider the conditions which might produce the
density in the shells around the radio bubbles in Abell~2052,
where the ratio of densities is about $\rho_f/\rho_a \sim 2$ (BSMW). 
The synchrotron age of the radio source in Abell~2052 is estimated to be
$t_1 \approx 9 \times 10^6$ yr (BSMW).
We first consider the possible role of cooling in creating the dense
shells.
The {\it Chandra} image and spectra imply that the present-day integrated
isobaric cooling time is $t_{\rm cool} \approx 2.6 \times 10^8$ yr (BSMW).
However, the theoretical expressions given above are given in terms of the
initial instantaneous cooling time $\tau_c$.
The integrated isobaric cooling time (the time to cool to zero temperature
at constant pressure) for bremsstrahlung cooling is $ 2 \tau_c / 3$.
This is related to the present-day integrated isobaric cooling time by
\begin{equation}
\frac{2}{3} \, \tau_c =
t_{\rm cool}
\left( \frac{P_i}{P_a} \right)^{1/2}
\left( \frac{\rho_f}{\rho_i} \right)^{3/2}
\, .
\label{eq:tcool_initial}
\end{equation}
Assuming maximal cooling, the cooling time which enters into
the cooling term (second term on r.h.s.) of equation~(\ref{eq:max_cool_f})
is the integrated isobaric cooling time after the shock compression,
which is
\begin{equation}
\tau_{\rm min} \equiv \frac{2}{3} \, \frac{\alpha^{1/2}}{\eta^{3/2}}
\, \tau_c 
= 
t_{\rm cool} 
\left( \frac{P_1}{P_a} \right)^{1/2}
\left( \frac{\rho_f}{\rho_1} \right)^{3/2} \, .
\label{eq:tcool_critical}
\end{equation}
Using equation~(\ref{eq:max_cool_f}), this gives
\begin{equation}
\tau_{\rm min} = t_{\rm cool} \, 
\left( \frac{P_1}{P_a} \right)^{-2/5} \,
\left( 1 -  \frac{t_1}{\tau_{\rm min}} \right)^{-1}
\, .
\label{eq:tcool_critical2}
\end{equation}
This equation can be solved to give
\begin{equation}
\tau_{\rm min} =
t_1 +
t_{\rm cool} \, \left( \frac{P_1}{P_a} \right)^{-2/5} \, .
\label{eq:tcool_critical3}
\end{equation}

Based on the estimated age of the radio source and the observed cooling
time in Abell~2052, the observed cooling time in the shells is much greater
than the age of the source, $ t_{\rm cool} \gg t_1$.
Thus, either $( P_1 / P_a ) \gg 1$ or
$\tau_{\rm min} \sim t_{\rm cool} \gg t_1$.
In the latter case, the cooling term in equation~(\ref{eq:max_cool_f})
(second term on r.h.s.) is very close to unity, and cooling is not
important.
Alternatively, if $( P_1 / P_a ) \gg 1$, then we can have
$t_{\rm cool} \sim t_1$, and cooling might be important.
This implies that the gas underwent a very strong shock, which greatly
increased its pressure, and has since undergone a very large adiabatic
expansion, which increased the cooling time to the large observed value.
For the observed values in Abell~2052, $ t_{\rm cool} \approx 30 t_1$.
Thus, for cooling to be important, we require that
$( P_1 / P_a )^{2/5} \ga 30$ or $( P_1 / P_a ) \ga 4000$.
This pressure ratio can also be written as
$( P_1 / P_a ) = ( P_1 / P_i ) ( P_i / P_a )$.
The second term represents the ratio of the ambient pressure at the initial
location of the gas divided by the ambient pressure at its present
location.
One expects the radial pressure gradient in a cooling flow to nearly balance
the gravitational potential;
if the shocked gas was lifted to a larger radius by the expansion of the
radio source or by buoyancy, then one expects $ P_i > P_a $.
However, unless the gas originated at a much smaller radius (which is
unlikely given the substantial mass of the gas in the shells in
Abell~2052, see BSMW, Blanton et al. 2002), one would only expect that $ P_i \la 2 P_a $
Thus, cooling can only be important if the shock increased the
pressure on the gas by a large factor,
$ ( P_1 / P_i ) \equiv \alpha \ga 2000$.
This requires a large shock Mach number, ${\cal M} \ga 40$.

To drive a shock with such a large Mach number through the large mass
of observed shell gas
($\sim 5 \times 10^{10} \, M_\odot$ in Abell~2052; Blanton et al. 2002),
the total energy of the ``explosion'' must be large
($\gtrsim 10^{62} \, {\rm erg}$).
If released in $<10^7 \, {\rm yr}$, it implies a power of 
$\gtrsim 10^{48} \, {\rm erg} \, {\rm s}^{-1}$, which we consider 
unrealistically high. 
In addition, as the shock propagates to outer lower density regions, it 
heats up these regions to very high temperatures. 
These lower density regions will not have time to cool, hence this scenario
predicts large regions with very high temperatures just outside the dense 
shells, contrary to the observations.
It would also require a coincidence to have the very strong initial shock
and the subsequent radiative cooling nearly cancel one another,
leaving the shell gas with densities and temperatures which are close
to the ambient values at present.
We therefore rule out the very high Mach number (strong shock) scenario.

If cooling is not important, then the high density of the shell material
must be the result of shock compression and adiabatic expansion.
For either equation~(\ref{eq:max_cool_f}) or (\ref{eq:min_cool_f}),
the first term on the r.h.s.\ is related to the entropy jump in the shock
to a power of $-3/5$, hence it is $\le 1$. 
For Mach numbers of ${\cal M}=2.5$, 5, and 10,
its value is 0.80, 0.46, and 0.21, respectively. 
For both equations, the third term on the r.h.s.\ is $\ge 1$,
since the bubble will expand and rise buoyantly outwards to regions of lower
density in the cooling flow.
However, it is possible that increasing the initial density of the gas
will also increase its initial pressure, which would make the fourth term,
which is $\le 1$, smaller. 
Note that the third and fourth terms are related to
the outward entropy jump to a power of $-3/5$.
Observed cooling flows always have convectively stable (i.e., increasing)
entropy gradients, and thus the combination of terms three and four is always
$\ge 1$.
For a very weak shock and a shallow inner pressure profile, the lifted
gas will have a density larger than its ambient value by a factor of
$\sim \rho_i/\rho_a$.
However, as mentioned above, to form a clear shell or filaments, we require
a shock, with, say, ${\cal M} = 5$.
This means that the first term on the r.h.s will be $\sim 1/2$. 
Hence, we require that in Abell~2052, the cool material had
an initial density ratio of $\rho_i/\rho_a \gtrsim 4$. 
For a density profile of $\rho \propto r^{-1}$, this implies that the
gas at $\sim 14$ kpc was lifted from $\sim 3.5$ kpc. 

One concern with this simple model is whether there was enough gas 
at smaller radii to form the observed shells.
Using the observed present day gas density distribution in Abell~2052
(equation~\ref{eq:rho_profile} below) and extrapolating into smaller radii 
than those observed, we find that the total mass 
of gas within 7 kpc of the cluster center, where the density is 
$\sim 4$ times higher than on the upper portion of the bubbles,  
is only about $10^{10}$ $M_\odot$,
which is smaller than the observed total mass of the two shells
in this cluster, $\sim 5 \times 10^{10} M_\odot$ (Blanton et al. 2002).
This discrepancy suggests that there are other factors which
have not been taken into account by our simple scenario.
One possibility is that the medium was highly inhomogeneous.
Inhomogeneity in the cooling flow medium is inferred from other arguments  
(Fabian 1994). 
We note in $\S 4$ below that the interface between the dense shells and the
surrounding ambient gas is unstable, and mixing is expected. 
 The mixing of cooler with hotter gas on small scales can lead to 
heat conduction.
The cooler gas is radiating more efficiently, cooling the hotter gas.
Therefore, cooling may be more important than what was assumed here for a 
homogeneous medium.  
This idea is supported by the strong correlation between the dense X-ray 
shells and the H$\alpha$ filaments (BSMW). 
Also, in an inhomogeneous medium the high density phase emits more efficiently.
Hence, the total mass expected in the shells can be overestimated. 
This by itself can't erase the discrepancy we find above, but it can reduce it. 
Another possibility is that the density profile in the inner regions
was much steeper before the inflation of the bubbles. 
This adds more available mass, and the high density shell could have
been lifted from farther out. 
 If, for example, the dense gas was
lifted from $\sim 10$ kpc, instead of from farther in, 
the mass within 10 kpc of the center would be then 
$\sim 3 \times 10^{10} M_\odot$.
Considering other uncertainties, a little steeper density
gradient before the inflation of the bubbles may account for the discrepancy. 
Another interesting possibility is that the age of the bubbles is
larger than the age inferred from the radio emission.
However, it can't be much longer than the buoyant rising time of
$\sim 2 \times 10^{7}$ yr (BSMW).  
It is quite plausible that a few of the factors discussed above act together,
accounting for the mass discrepancy between the observed shells' mass 
and the estimate from our simple model. 

Finally, we note that the derivation of equation (\ref{eq:max_cool_f})
assumes pressure equilibrium between the dense, X-ray bright shell 
and the ambient gas. 
Therefore, a density ratio of $\sim 2$ implies a temperature 
ratio of $\sim 0.5$, meaning that the mildly shocked gas (${\cal M} \sim 5$)
is cooler than the ambient gas, as observed. 

\section{Formation of a Bubble} \label{sec:bubbles}

In this section we estimate the size of the bubble inflated
by the injection of energy from the central AGN.
Recent numerical simulations of bubble evolution can be found in
Churazov et al.\ (2001),  Br\"uggen et al.\ (2002),  
Quilis et al.\ (2001), Saxton et al.\ (2001),  and
Br\"uggen \& Kaiser (2001), while relevant recent numerical simulations of 
jets were conducted by, e.g., Clarke, Harris \& Carilli (1997), 
Rizza et al.\ (2000), and Reynolds et al.\ (2001).  
The general formation of bubbles from AGN was studied previously via
numerical simulations, e.g., Wiita (1978) and Norman et al.\  (1981),
and a detailed 
analytical study of bubble and jet formation was conducted
by Smith et al.\ (1983).
Here, we provide approximate analytic expressions.
Although these may be less accurate in any individual case than
specific numerical simulations of that system, they are more
general and can be applied more easily to new observations.
 
Although the derivations in this section are general, we scale
quantities according to the cooling flow cluster Abell~2052.
For the radial variation of the gas pressure and density in Abell~2052,
we approximate the results of BSMW (Fig.~2) by
\begin{equation}
P(r) \approx 1.7 \times 10^{-10}
\left( \frac{r}{30 \, {\rm kpc}} \right)^{-1.0}
\, {\rm dyn} \, {\rm cm}^{-2}
\qquad r > 30 \, {\rm kpc}
\, ,
\label{eq:P_profile}
\end{equation}
and
\begin{equation}
\rho(r) \approx 4.2 \times 10^{-26}
\left( \frac{r}{30 \, {\rm kpc}} \right)^{-1.1} 
\, {\rm g} \, {\rm cm}^{-3}
\qquad r > 30 \, {\rm kpc}
\, ,
\label{eq:rho_profile}
\end{equation}
where $r$ is the distance from the cluster center.
Within 30 kpc, the observed gas structure in Abell~2052 is strongly affected
by the bubbles.
However, we assume that the same profiles also applied to the interior
regions of the cooling flow prior to its disruption by the AGN.

A lower limit on the energy blown into a bubble of radius $R_b$ is the
work which it has done on the surrounding medium,
$E_{min}=[(\gamma/(\gamma-1)](4 \pi /3)R_b^3 P_e$,
where $P_e$ is the external pressure.
Accounting both for the work done by the bubble and its present internal
energy, will give $\sim 2$ times as much energy. 
We will assume that the main pressure support and internal energy in the
bubble is due to very hot non-relativistic thermal gas, rather than magnetic
fields or relativistic particles.
In Abell~2052 and Perseus, the minimum radio pressure in the bubble
is about a factor of ten smaller than the external pressure
(Fabian et al.\ 2000; BSMW).
This suggests that magnetic fields and relativistic electrons may not
be the main source of pressure.
Taking the center of the bubble at $r_{\rm kpc}=14$ we crudely estimate
$E_b \simeq 3 \times 10^{59}$ erg.

For simplicity, we will divide the evolution of the bubble into two phases:
an energy injection phase lasting a time of $\tau_I$, and a later phase
when the energy injection rate is much lower and is assumed to be zero.
Let $t$ be the total lifetime of the bubble from the start of the energy
injection phase.
We will assume that the bubble undergoes spherical expansion into
a constant density medium.
We will take this constant density $\rho_c$ to be the value in the undisturbed
cooling flow at the distance $r_c$ from the center of the cooling flow
to the center of the bubble, $\rho_c = \rho ( r_c )$.
This might be a good approximation if the radius of the bubble $R_b$
were a small fraction of $r_c$, and if the bubble were not buoyant.
Most of the observed bubbles have $ R_b \sim r_c $.
As a result, the gas density in the ambient medium should vary with
position around the bubble.
Also, the bubble will both expand and rise buoyantly outward, and
this also should cause a variation in the ambient density around the
bubble.
As the bubble rises and expands, it will lift dense material from the very
inner regions of the cooling flow ($\sim 1$ kpc);
the shell will not consist solely of material originating from its
present-day center at $r_c \sim 14$ kpc. 
The other major approximation we make is to ignore the thermal pressure of
the ambient medium exterior to the bubble. 
This may be a good assumption at early stages, but not at late stages 
when the Mach number of expansion of the bubble is small. 
Small Mach numbers at the present time are implied by the lack of clear shock
structures surrounding the observed bubbles
(Fabian et al.\ 2000; BSMW).

We use the expression given by Castor, McCray \& Weaver (1975) for the
expansion of an interstellar bubble;
Bicknell \& Begelman (1996) also applied this expression to study the
bubble formed by the radio jet in M87.
Scaling to a constant density $\rho_c$ appropriate for $r_c \sim 5$ kpc
in Abell~2052,
the radius of the bubble under these assumptions is given by 
\begin{equation}
R_b \simeq 7.8 
\left(\frac {t}{10^7 \, {\rm yr}} \right)^{3/5} 
\left(\frac {\dot E}{10^{44} \, {\rm erg} \, {\rm s}^{-1}} \right)^{1/5} 
\left(\frac {\rho_c}{10^{-25} \, {\rm g} \, {\rm cm}^{-3}} \right)^{-1/5}
\, {\rm kpc} \, .
\label{eq:bubble_radius1}
\end{equation}
The energy injection rate used here is similar to that determined for the
Perseus cluster, where Fabian et al.\ (2002) constrain the jet power 
to be between $10^{44}$ and $10^{45} \, {\rm erg} \, {\rm s}^{-1}$.
The calculation above neglects the effect of the ambient pressure on the
bubble, therefore it overestimates the radius at late evolutionary stages. 
Comparing with the results of Br\"uggen et al.\ (2002), we
find that we overestimate the radius of the bubble by 70\%
for the case $\dot E = 4.4 \times 10^{41} \, {\rm erg} \, {\rm s}^{-1}$ at 
$t=8.36 \times 10^6$ yr (their Fig.~2), and by 45\% for the case 
$\dot E = 3.8 \times 10^{42} \, {\rm erg} \, {\rm s}^{-1}$ at
$t=1.25 \times 10^7$ yr (their Fig.~4).
For the energetic case of $\dot E = 10^{44} \, {\rm erg} \, {\rm s}^{-1}$ at 
$t=5 \times 10^6$ yr, on the other hand, equation~(\ref{eq:bubble_radius1})
gives the same radius as their simulation (their Fig.~5).    
Thus, equation~(\ref{eq:bubble_radius1}) is a good approximation in the
expected limit.

In the subsequent expressions for the bubble evolution,
it is useful to replace the energy deposition rate $\dot{E}$ with
$E_b / \tau_I$, and to replace the age of the bubble $t$ with
its radius $R_b$.
The expansion velocity of the bubble surface as a function of its radius
is given by 
\begin{equation}
v_b (R_b) \simeq 570  
\left(\frac {\tau_I}{10^7 \, {\rm yr}} \right)^{-1/3} 
\left(\frac {E_b}{10^{59} \, {\rm erg} } \right)^{1/3} 
\left(\frac {\rho_c}{10^{-25} \, {\rm g} \, {\rm cm}^{-3}} \right)^{-1/3} 
\left(\frac {R_b}{10 \, {\rm kpc}} \right)^{-2/3} 
\, {\rm km} \, {\rm s}^{-1} \, .
\label{eq:bubble_velocity1}
\end{equation}
 At the end of the injection phase the bubble radius is 
\begin{equation}
R_b (\tau_I) \simeq 9.8 
\left(\frac {\tau_I}{10^7 \, {\rm yr}} \right)^{2/5} 
\left(\frac {E_b}{10^{59} \, {\rm erg} } \right)^{1/5} 
\left(\frac {\rho_c}{10^{-25} \, {\rm g} \, {\rm cm}^{-3}} \right)^{-1/5} \, {\rm kpc} \, ,
\label{eq:bubble_radius2}
\end{equation}
and its surface expansion velocity at the end of the energy injection phase is 
\begin{equation}
v_b (\tau_I ) \simeq 580
\left(\frac {\tau_I}{10^7 \, {\rm yr}} \right)^{-3/5} 
\left(\frac {E_b}{10^{59} \, {\rm erg} } \right)^{1/5} 
\left(\frac {\rho_c}{10^{-25} \, {\rm g} \, {\rm cm}^{-3}} \right)^{-1/5} \, {\rm km} \, {\rm s}^{-1} \, .
\label{eq:bubble_velocity2}
\end{equation}
The expressions we use for the bubble expansion require that it be highly
supersonic, which implies $v_s \gg c_s \simeq 820 \, {\rm km} \, {\rm s}^{-1}$
for a temperature of $3 \times 10^7$ K. 
Thus, the expressions we give will become inaccurate when $R_b \ga 10$
kpc.

\section{Rayleigh-Taylor Instability and Bubbles}
\label{sec:rt}

The deceleration of the bubble surface for $t \leq \tau_I$ is given by  
$a_b \equiv \dot v_b = - (6/25) R_b t^{-2}$. 
Written as a function of time (for $t \leq \tau_I$), the deceleration is
\begin{equation}
a_b \simeq -  7.3 \times 10^{-8}
\left(\frac {\tau_I}{10^7 \, {\rm yr}} \right)^{-1/5} 
\left(\frac {E_b}{10^{59} \, {\rm erg} } \right)^{1/5} 
\left(\frac {\rho_c}{10^{-25} \, {\rm g} \, {\rm cm}^{-3}} \right)^{-1/5} 
\left(\frac {t}{10^7 \, {\rm yr}} \right)^{-7/5} \, {\rm cm} \, {\rm s}^{-2}
\, ,
\label{eq:deceler1}
\end{equation}
while as a function of the bubble radius, it is
\begin{equation}
a_b \simeq -  7.7 \times 10^{-8}
\left(\frac {\tau_I}{10^7 \, {\rm yr}} \right)^{-2/3} 
\left(\frac {E_b}{10^{59} \, {\rm erg} } \right)^{2/3} 
\left(\frac {\rho_c}{10^{-25} \, {\rm g} \, {\rm cm}^{-3}} \right)^{-2/3} 
\left(\frac {R_b}{10 \, {\rm kpc}} \right)^{-7/3}  \, {\rm cm} \, {\rm s}^{-2}
\, .
\label{eq:deceler2}
\end{equation}

This deceleration (or negative acceleration) is equivalent to a
gravitational acceleration in the outward radial direction (or a positive
gravitational acceleration $g$).
We determine the gravitational acceleration from the
assumption of hydrostatic equilibrium,
\begin{equation}
g = \frac{1}{\rho(r)} \, \frac{ d P(r)}{d r} \, .
\label{eq:hydrostatic1}
\end{equation}
For the pressure and density profiles for the gas in Abell~2052
(equations~\ref{eq:P_profile} and \ref{eq:rho_profile}),
the acceleration is
\begin{equation}
g \simeq - \frac{P}{\rho r} \simeq 
- 1.2 \times 10^{-7} 
\left(\frac {r}{10 \, {\rm kpc}} \right)^{-0.9}  \, {\rm cm} \, {\rm s}^{-2}
\, .
\label{eq:hydrostatic2}
\end{equation}
For bubbles inflated about a center which is not the cluster center,
the least stable portion of the bubble is likely to be its outermost
region.
Here, the dense shell lies above the low-density interior of the bubble.
Let $r_c$ be the radius of the center of the bubble;
this might correspond to the point at which a jet from the
center was stopped in a termination shock.
As before, $R_b$ is the radius of the bubble (measured from $r_c$).
Then the radius of the outermost part of the bubble is
$r=r_c+R_b$.
If we use equation~(\ref{eq:rho_profile}) for the density profile in
Abell~2052,
we find for
\begin{equation}
r =r_c+R_b \la 15 
\left(\frac {\tau_I}{10^7 \, {\rm yr}} \right)^{0.4} 
\left(\frac {E_b}{10^{59} \, {\rm erg} } \right)^{-0.4} 
\left(\frac {R_b}{10 \, {\rm kpc}} \right)^{1.4}
\, {\rm kpc} \, ,
\label{eq:rt1}
\end{equation}
the magnitude of the gravitational acceleration exceeds that of the
deceleration, and the interaction between the low density bubble and
the dense, X-ray bright shell is Rayleigh-Taylor (RT) unstable. 
With the scaling used above, we find that the expansion of the
bubble is RT stable as long as the bubble radius is 
$R_b \la 4$, 6, or 8 kpc, for $r_c = 0$, 1, and 3 kpc, respectively.
For a larger energy $E_b$, the shell remains stable to later times;
for example,
for $ E_b = 2 \times 10^{59}$ erg and $r_c=0$,  
the shell is stable as long as $R_b \la 8$ kpc. 
At later stages, the shell becomes unstable. 
However, in our treatment of the bubble expansion, we neglected the ambient
pressure (we assumed highly supersonic expansion).
At late stages, the ambient pressure will increase the
deceleration of the shell, making it more stable against RT modes.
We conclude that the shell becomes RT-unstable only when it is large, and
only on its outermost segment.
Along other segments, the gravitational acceleration component perpendicular 
to the interface is lower, and the expansion is RT-stable until the 
bubble almost stops expanding.  
 
The growth time for RT instabilities (the e-folding  time) is given
by $\tau_{\rm RT} = | g_e k |^{-1/2}$,
where $\lambda$ and $k=2 \pi /\lambda$ are the wavelength and wavenumber of
the mode being considered. 
The effective gravitational acceleration, $g_e$, at the outer part of the
shell is $g_e=|g|-|a_b|$.
For disruption of the entire bubble, we will consider a large scale mode
with $\lambda = 2 R_b$.
The time scale for the growth of this mode is
\begin{equation}
\tau_{\rm RT} = 1 \times 10^{7} 
\left(\frac {R_b}{10 \, {\rm kpc}} \right)^{1/2} 
\left(\frac {g_e}{1 \times 10^{-7} \, {\rm cm} \, {\rm s}^{-2}} \right)^{-1/2}
\, {\rm yr} \, .
\end{equation}

The growth time for RT instabilities as given above, is comparable
to the estimated age of the bubbles;
for example, in Abell~2052 the estimated age of the radio bubbles is
about $9 \times 10^6$ yr (BSMW).
This suggests that in most cases the bubble will be stable during its
inflation phase.
In other cases the bubbles will start to be fragmented, but only in the
outer regions.
This is in reasonable agreement with the fact that the bubbles in Abell~2052
are fairly complete, except for possible gaps at their outer edges 
(the north side for the north bubble and the south side for the south bubble).

At the outer edge of the dense shell, there is a significant drop in the 
density, and this interface would be unstable when the inner 
discontinuity is stable.
However, at early times when the bubbles are small, they are expected to
expand supersonically (equation~\ref{eq:bubble_velocity1}), and the outer
density discontinuity is the result of a shock rather than
merely a contact discontinuity.

\section{Ghost Bubbles}
\label{sec:ghost}

There are several cases of outer, radio faint, isolated bubbles, 
or ``ghost bubbles,''
including those in the Perseus cluster (Fabian et al.\ 2000, 2002),
Abell~2597 (McNamara et al.\ 2001),
and possibly MKW3s (Mazzotta et al.\ 2001).
These bubbles are believed to have risen outward buoyantly.
The bubbles are X-ray faint, and no trail of bright X-ray material is 
observed behind them. 
The two outer bubbles in Perseus are isolated and have well defined 
shapes, which are elongated in the azimuthal direction. 
For the northwestern and southern bubbles, the distances from the center
of the cluster to the center of the bubbles are
1.5 times and 2 times their widths, respectively. 
We assume that the line connecting the bubble centers to the center of
the cluster is inclined by $\sim 45^\circ$ to our line of sight, so that
the bubbles are actually at distances which are $\sim 3$ times
their diameter.
  
Churazov et al.\ (2001) give the rising velocity of a buoyant
spherical bubble as $ v_r \simeq (8 g R_b /3C)^{1/2}$, where the drag
coefficient is $C \simeq 0.75$. 
Multiplying this velocity by the RT e-folding time, we find that the 
distance $D$ a bubble will be buoyant before starting to be 
disrupted in an observationally noticeable way is $r \simeq R_b$.
Hence, a bubble will not rise much before being disrupted by RT instability
modes.
If the bubbles in Perseus were inflated to their diameters, as in Abell~2052,
and then rose outward due to buoyancy to a total distance of $\sim 3$ times
their diameter,
their age must be $\sim 4-5$ times the growth time of large-scale RT modes.  
Thus, we would expect these bubbles to have undergone considerable
disruption.
Indeed, carefully examining the available images of the pair of outer 
bubbles in Perseus (Fabian et al.\ 2000, 2002) we identify what seems to 
be a RT-protrusion in the northwest bubble, and some other irregularities 
in the southern bubble.
We attribute these to developed RT instability modes.
We also attribute the
elongation of the bubbles in the azimuthal direction to
RT instabilities;
this cannot be a result of the bubble reaching a radius at which its density
equals the ambient density, since in that case the interior of the bubbles
would not be X-ray faint.

The presence of bubbles even further from the centers of clusters can be
explained in two ways.
First, magnetic fields may suppress RT instabilities.
Fabian et al.\ (2002) argued that magnetic suppression of RT instabilities
was needed to explain the sharp edges in the outer bubbles in Perseus.
However, the suppression is efficient only for short wavelengths, and not
for large wavelength modes that disrupt the bubble.
Second, it is possible that the bubbles were formed farther from the
center, by a jet expanding outward.  
An example of this process in action might be Cygnus-A
(Smith et al.\ 2002).
However, Cygnus-A is somewhat unusual, as it is the most luminous radio
source in the nearly universe.
In most cases, radio jets don't propagate to large distances in
cooling flow clusters.
Thus, we don't expect to find large bubbles
beyond a radius of $\sim 50$ kpc. 

\section{Interaction of Bubbles} \label{sec:interaction}
  
The existence of ghost bubbles in clusters with more centrally located,
radio bright bubbles or lobes
(Fabian et al.\ 2000, 2002; McNamara et al.\ 2001) suggests
that the radio sources at the centers of cooling flows are episodic.
Thus, we now consider the interaction of a new radio bubble with a
previously inflated ghost bubble.
Assume that a currently expanding bubble runs into a small, low-density cavity 
due to a ghost bubble.
A small portion of solid angle $\Omega \ll 4 \pi$ of the bubble's 
surface is assumed to enter the cavity (ghost bubble).
We will call the portion of the dense bubble's shell that enters the cavity
a ``blob.''
We further assume that the ghost bubble is elongated in the radial direction,
such that the cross section stays constant;
such a bubble is not like the bubbles observed in Perseus, but more like
one left by a radio jet.
We also neglect the pressure inside the cavity, and since we assume
that the volume of the cavity is small, we neglect the decrease in
the thermal pressure inside the current bubble due to the expansion of
its thermal gas to the volume of the cavity. 
The effects we neglect here reduce the efficiency of the acceleration 
process.
However, later we neglect effects which increase the efficiency of
the proposed scenario.

At the collision time $t_c$, measured from the birth of the active bubble,
the bubble radius is $R_c$. 
As in \S~\ref{sec:bubbles},
we assume a constant energy injection rate into the bubble, 
and the bubble expands inside a medium of constant density $\rho_c$. 
The radius of the bubble as a function of time is given by
equation~(\ref{eq:bubble_radius1}).
The pressure $P$ just inside the bubble acts on its surface, such that the
momentum changes according to $d (M v_b)/dt=4 \pi R_b^2 P$,
where $M=4 \pi R_b^3 \rho_c /3$ is the mass of the dense 
shell on the surface of the bubble,
and $v_b= dR_b/dt=(3/5)R_b/t$ is its expansion velocity.
The portion of the bubble's surface that enters into the low-density 
cavity has a mass $M_{\rm bl} = (\Omega/4 \pi)M$,
which stays constant for times $t > t_c$.
The rate of change of the momentum of that mass is    
$M_{\rm bl} (dv_{\rm bl}/dt) = \Omega R_b^2 P$.  
 From the expression for the rate of change in momentum for the bubble and
the blob (the portion running into the cavity) we find 
\begin{equation}
M_{\rm bl} \frac{dv_{\rm bl}}{dt} = \frac{\Omega}{4\pi} \frac{d( M v_b)}{dt} \, .
\label{eq:blob_momentum}
\end{equation}
 From the expressions for $M$ and $R_b$, we find that 
$dM/dt=4 \pi R_b^2 v_b \rho_c$, and $dv_b/dt=-(6/25)R_b/t^2$. 
Substituting these and the expression for $M_{\rm bl}$ in
equation~(\ref{eq:blob_momentum}),
we get
\begin{equation}
\frac{d v_{\rm bl}}{dt} = \frac{21}{25} \frac{R_b}{t^2} \, ,
\label{eq:blob_accel}
\end{equation}
with the initial condition $v_{\rm bl}=v_c=(3/5)R_c/t_c$ at $t=t_c$.
Using the dependence of $R_b$ on $t$, the last equation can be integrated
analytically to give
\begin{equation}
v_{\rm bl} = \frac {9}{2} v_c \left[ 1- \frac{7}{9} 
\left( \frac{t}{t_c} \right)^{-2/5} \right] \qquad t \ge t_c \, ,
\label{eq:blob_vel}
\end{equation}
where $v_c=(3/5)R_c/t_c$ is the velocity of the shell and blob at the time 
of the collision, $t=t_c$. 
The last equation can be integrated analytically to give the
distance of the blob from the center of the bubble 
\begin{equation}
\frac{R_{\rm bl}}{R_c}=\frac{9}{5}
-\frac {7}{2} \left( \frac{t}{t_c}\right)^{3/5}
+\frac{27}{10}\frac {t}{t_c} \qquad t \ge t_c \, .
\label{eq:blob_radius}
\end{equation}
The idealized picture that the bubble continues to expand
undisturbed breaks down as evolution proceeds. 
But for a time $t \sim 2 t_c$, we can assume it is adequate, and  
find $v_{\rm bl}=1.85 v_c=2.4 v_b$, where $v_b$ is the expansion
speed of the bubble surface at $t=2t_c$. 
During that time, the bubble has expanded by a factor of $2^{3/5}=1.52$,
i.e., $R_b=1.52 R_c$, while the blob is at a distance of 
$R_{\rm bl}=1.89R_c=1.25 R_b$ from the center of the bubble. 
The blob is accelerated quite efficiently, and can reach a relatively
high velocity.  
If the bubble surface velocity was originally $v_c=400$ km s$^{-1}$,
then the blob can have $v_{\rm bl} \sim 800$ km s$^{-1}$.  

\begin{figure}
\plotone{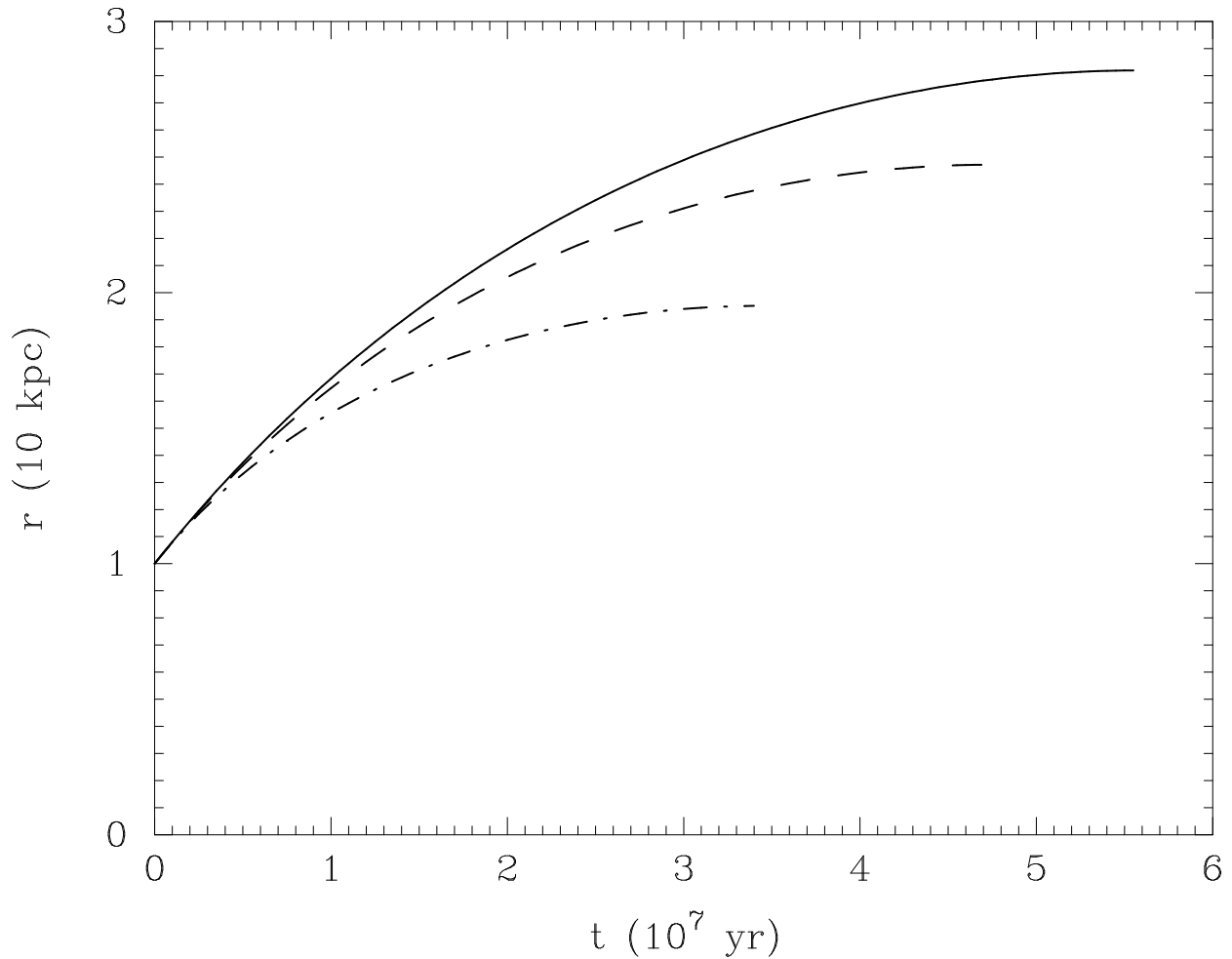}
\caption{
The location of a quickly outward moving dense blob as a function of time.
Plotted are solutions of
equation~(\protect\ref{eq:blob_motion})
with initial velocity of 800 km s$^{-1}$ at $r=10$ kpc, and
$\rho_c/l= 0.06 \rho _{\rm bl}$/kpc
(e.g., $l=5$ kpc, $\rho_c = 0.3 \rho_{\rm bl}$)
(solid line); 
$\rho_c/l= 0.1 \rho_{\rm bl}$/kpc (dashed line);
$\rho_c/l= 0.23 \rho_{\rm bl}/ \, {\rm kpc}$ (dashed-dotted line).
Here, $l$ is the length of the blob and $\rho_c$ and $\rho_{\rm bl}$
are the ambient and blob densities, respectively. 
\label{fig:blob}
}
\end{figure}

Eventually, the accelerated blob will leave the cavity, enter
a high density region, and be decelerated due to the drag with its 
surroundings and gravity. 
 We neglect the buoyant force, and assume that the acceleration by the
high-pressure bubble interior ceased; both effects, if included, will lower 
the deceleration and make the blob reach larger distances. 
 From Churazov et al.\ (2001) we find the gravitational acceleration in M87
to be 
$g=-5 \times 10^{-8} (r/10 \, {\rm kpc})^{-1/2} \, {\rm cm} \, {\rm s}^{-2}$, 
where $r$ is the distance from the
cluster center.
The drag force can be written as $F_D=0.5 C S v_{bl}^2 \rho_a$,
where $C \simeq 0.75$ is the drag coefficient, $S$ the cross section of the
blob, $v_{\rm bl}$ its velocity, and $\rho_a$ the ambient density.
We will treat the blob as a cylinder of length $l$ and radius $R$
whose axis of symmetry is parallel to the direction of motion.
Its density is $\rho_{\rm bl}$, so that its mass is $M_{\rm bl}=\rho_{\rm bl} \pi R^2 l$. 
The drag deceleration is then
$a_D=-F_d/M_{\rm bl} \simeq 0.4(v^2/l)(\rho_a/\rho_{\rm bl})$.
Scaling the variables, the equation of motion for the blob becomes
\begin{equation}
\frac{d^2 r}{dt^2} \simeq
-5 \times 10^{-8} \left [
\left(\frac {r}{10 \, {\rm kpc}} \right)^{-1/2} +
\left(\frac {l}{5 \, {\rm kpc}} \right)^{-1} 
\left(\frac {v}{800 \, {\rm km} \, {\rm s}^{-1}} \right)^{2} 
\left(\frac {\rho_a}{0.3 \rho_{\rm bl}} \right) \right] \, {\rm cm} \, {\rm s}^{-2} \, .
\label{eq:blob_motion}
\end{equation}
We solved this equation numerically for a blob leaving the ghost bubble cavity
and entering the high density medium at $r=10$ kpc.
The outward velocity of the blob is 800 km s$^{-1}$ at $r= 10$ kpc.
In Figure~\ref{fig:blob}, we plot $r$ as function of time for three sets of
parameters.
The figure shows that a blob can reach relatively large
distances from the cluster center in reasonably short times.

We propose that such a blob, produced by the interaction between bubbles,
may explain the X-ray and radio structure in the eastern radio `ear' in M87
(Belsole et al.\ 2001; Churazov et al.\ 2001 and references therein).
Unlike bubbles in other cooling flow clusters, there is a positive
correlation between X-ray and radio emission on the long jet-like
feature connecting the center of M87 with the eastern radio `ear' (lobe)
(e.g., Churazov et al.\ 2001). 
Churazov et al.\ (2001) suggest that the eastern radio lobe of M87 
is a buoyant bubble which drags dense material with it, giving the positive
correlation between the X-ray and radio emission. 
Saxton et al.\ (2001) suggest a similar explanation
for the northern middle lobe in the moderate cooling 
flow cluster (Allen et al.\ 2001a) Centaurus A (Abell~3526). 
Both these papers present gas-dynamical simulations of buoyant bubbles.
Their model seems to be unable to explain all cases, e.g.,
the SW lobe in Hydra A (N2002). 
That model may have problems even in M87, its prime target. 
The general X-ray structure of M87 as revealed in {\it Chandra} archive 
data (see also XMM-Newton observations by Belsole et al.\ 2001), 
does not resemble the structure Churazov et al.\ (2001)
find in their numerical simulations. 
Indeed, in a recent paper, Br\"uggen et al.\ (2001) find
in their numerical simulations that the X-ray emissivity along the simulated
jet is lower than the ambient emissivity, which contradicts the 
explanation of Churazov et al.\ (2001) and Saxton et al.\ (2001).
We here propose a tentative scenario, where the dense X-ray 
material was ejected from the central region of M87 as a bullet.
The jet, which formed the ear, formed the cavity through which 
dense X-ray material (the bullet) expanded outward.  
The distance of the ear from the cluster center is $\sim 20$ kpc,
which can easily be reached by the bullet (see fig. 1). 
 The idealized bullet, in the calculation presented here, is a stream of
dense X-ray material pushed by an inner bubble. 
 Inspecting {\it Chandra} images, we find hints for a few low X-ray
emissivity bubbles near the base of the jet-like feature.
Careful analysis of the images of M87, and numerical simulations of bubble
interactions are required to disprove or support our tentative scenario.

\section{Discussion and Summary} \label{sec:discussion}

We derived simple expressions which can be used to analyze some
properties of X-ray deficient bubbles in clusters of galaxies. 
We demonstrated their application for Abell~2052 and Perseus.
We also addressed the X-ray and radio structure to the east of 
the center of M87. 
 
Our main results can be summarized as follows:

(1) The cool material which forms the X-ray bright shells around
the X-ray deficient radio bubbles in the cooling flow cluster Abell~2052,
and very likely in other cooling flow clusters (e.g., Perseus: Fabian et al.
2002, Hydra A: N2002),
was lifted from the very inner regions of the cooling flow, 
rather than being shocked and then cooled substantially due to shock 
compression.
The material went through a shock, but a mild one, the
main effect of which was to shape the material in a dense shell around
the radio bubble.
To achieve the present shell-to-ambient density ratio of $\sim 2$,
the dense gas had to have an initial density $\sim 4$ times 
its present ambient density.
This is quite plausible for dense gas lifted from $\sim 5-10$ kpc. 
One problem with this model is that the observed mass in the shells
in Abell~2052 is somewhat larger than the expected mass in these inner 
regions,
determined by extrapolating the observed gas density profile into the center.
We suggest that a steeper density gradient in the inner region prior 
to the bubble inflation, and/or inhomogeneities of the shell and ambient
medium can account for this mass discrepancy in our simple model.

(2) The presence of cool material in the X-ray shells surrounding
the bubbles has implications for a possible strong AGN shock propagating 
from the cluster center at the onset of the radio activity.  
Such a shock is not observed presently in any cluster,
but may have a duty cycle on a long time scale.
Soker et al.\ (2001) proposed that such strong intermittent
shocks, with shock velocities $\gtrsim 7000$ km s$^{-1}$,
may result in much lower mass cooling rates. 
If, for example, 
the presently cool shell was at an initial 
temperature of $\sim 10^{7}$ K, and then was hit by a strong shock 
with a speed of $\sim 7000$ km s$^{-1}$ (a Mach number ${\cal{M}} \sim 15$). 
 From equations~(\ref{eq:shock compression}), (\ref{eq:shock pressure}),
and (\ref{eq:max_cool_f}),
we find that the gas was shocked to a temperature of 
$\sim 6 \times 10^8$ K, and then cooled adiabatically
(the radiative cooling will be negligible during a time of 
$< 10^{8} \, {\rm yr}$) to $\sim 7 \times 10^7$ K.
This is still too hot to match the presently observed dense shell.
We therefore rule out such a scenario.
Cases with even higher Mach numbers were ruled out in $\S 2$.
We conclude that the presence of massive dense and cool shells
preclude that the radio activity was initiated by a very strong shock
(if the age of the present radio activity is $\lesssim 10^8$ yr).
However, this does not rule out a strong shock $\gtrsim 10^9$ yr ago.  
   
(3) From observations, it appears that the inflation of bubbles by 
AGN activity
in cooling flow clusters is a general phenomenon. 
Our estimate via simple approach of the bubble size explains this as
a result of the insensitivity of the bubble size
(equations~\ref{eq:bubble_radius1} and \ref{eq:bubble_radius2})
to the injection energy rate and ambient density.
Therefore, a bubble will be formed with a size of several kpc for
most reasonable levels of AGN activity. 

(4) During the early inflated phase, i.e., while energy is injected by the 
AGN, the interface between the hot bubble and the dense shell is 
Rayleigh-Taylor (RT) stable, explaining the smooth-surface bubbles
close to the center of cooling flow clusters (e.g., Perseus). 
Later, the upper segment of the shell becomes unstable, and the shell
may break-up there, e.g., the northern shell in Abell~2052. 
The interface between the dense shell and the outer lower density medium 
is RT-unstable, but the density ratio is close to unity, and considering
some stress and friction, the instability evolves slowly. 
It is quite possible that the dense shell will start to break-up by
these later RT-unstable modes.
After being buoyant to about twice their diameter, the interface between the
low density bubbles and the outer regions becomes RT-unstable, and
large RT-fingers with sizes not much smaller than the bubble size
may be observed penetrating the bubbles. 
The outer bubbles in Perseus seem to be disrupted by RT instabilities. 
   
(5) We study the case of a bubble expanding into an older bubble. 
The portion of the still-active bubble which enters the low density
cavity formed by the older bubble will be accelerated to higher
velocities, up to $\sim 2$ times the original expansion
velocity of the bubble. 
These quickly moving blobs may travel out to radii of $20-30$ kpc
in reasonable times.
We speculate that such high density blobs may explain the high 
density X-ray emitting regions along the radio structures in M87. 
In M87, the cavity was formed by a radio jet which formed
the eastern `ear', and the blobs created a stream of dense X-ray emitting gas.

\acknowledgements
N. S. was supported by a Celerity Foundation Distinguished Visiting 
Scholar grant at the University of Virginia, and by a grant from the
US-Israel Binational Science Foundation.
C. L. S. and E. L. B. were supported in part by the National Aeronautics
and Space Administration through {\it Chandra} Award Numbers
GO0-1158X
and
GO1-2133X,
issued by
the {\it Chandra} X-ray Observatory Center, which is operated by the
Smithsonian
Astrophysical Observatory for and on behalf of NASA under contract
NAS8-39073.
E. L. B. was supported by the {\it Chandra} Fellowship Program, through NASA grant
\#PF1-20017.

\end{document}